\documentclass{PoS}

\title{STACEX: RPC-based detector for a multi-messenger observatory in the Southern Hemisphere}

\ShortTitle{STACEX observatory}

\author{\speaker{Di Sciascio Giuseppe}\\
        INFN - Roma Tor Vergata, Via della Ricerca Scientifica 1, Roma, Italy\\
        E-mail: \email{giuseppe.disciascio@roma2.infn.it}}
 \author{Camarri Paolo, Santonico Rinaldo\\
        Dipartimento di Fisica Universit\'a di Roma Tor Vergata and INFN - Roma Tor Vergata, Italy
        }
\author{Cardillo Martina, Marchese Fabrizio, Piano Giovanni, Tavani Marco\\
        IAPS-INAF, Via del Fosso del Cavaliere 100, 00133, Roma, Italy }
\author{Bigongiari Ciro\\
        INAF-OAR, Via Frascati 33, 00078 Monteporzio Catone (RM), Italy
        }
\author{Bulgarelli Andrea, Fioretti Valentina\\
        INAF-OAS Bologna, via Gobetti 101, I-40129 Bologna, Italy. 
        }
 \author{Casanova Sabrina\\
       Institute of Nuclear Physics ul. Radzikowskiego 152, 31-342 Kraków, Poland 
        }
 
 \abstract{Extensice Air Shower (EAS) arrays are survey instruments able to monitor continuously all the overhead sky.
Their wide field of view (about 2 sr) is ideal to complement directional detectors by performing unbiased sky surveys, by monitoring variable or flaring sources, such as AGNs, and to discover transients or explosive events (GRBs).
With an energy threshold in the 100 GeV range EAS arrays are transient factories.
All EAS arrays presently in operation or under installation are located in the Northern hemisphere. A new survey instrument located in the Southern Hemisphere should be a high priority to monitor the Inner Galaxy and the Galactic Center.\\
STACEX is the proposal of a hybrid detector with ARGO-like RPCs coupled to Water Cherenkov Detectors (WCDs) mainly to lower the energy threshold at 100 GeV level.\\
In this contribution we introduce the possibility of improving the low energy sensitivity of survey instruments by equipping RPCs, which were proved to be optimal detectors at 100 GeV energies by the ARGO-YBJ Collaboration, with WCDs.
An EAS detector with high sensitivity between 100 GeV and 1 TeV would be a valuable complementary transient detector in the CTA era.
}

\FullConference{36th International Cosmic Ray Conference -ICRC2019-\\
		July 24th - August 1st, 2019\\
		Madison, WI, U.S.A.}

\begin{document}

\section{Introduction}

Gamma-ray astronomy is the observation of the cosmos through the detection of the highest-energy electromagnetic radiation, produced by the most energetic objects in the Universe and in the most extreme environments in terms of radiation, magnetic and gravitational fields. Many of the astrophysical objects known to be emitters of high-energy gamma rays within our Galaxy and outside the Milky Way burst or flare up in a unpredictable way over a wide range of timescales from milliseconds to years. The flares and the bursts are often associated to events happening in neutron stars and black holes and in the active nuclei of Galaxies and produce potential multi-messenger signatures of such extraordinary events including very high-energy cosmic rays, neutrinos and/or gravitational waves (GWs). Flaring and bursting sources such as AGNs or GRBs, which are so bright events to outshine all other astrophysical objects when they burst, are also the most likely candidate sources for the origin of cosmic rays up to EeV energies. While AGNs have been observed in the TeV band, many questions surround the nature of TeV emission. For example, how does the TeV emission differ during the AGN quiescent state compared to the flaring state ? What fraction of AGNs and GRBs emit GeV or TeV gamma rays ? How much energy is available in these sources to accelerate cosmic rays ?

Currently two different experimental techniques are applied in gamma-ray astronomy from ground: (1) Imaging Atmospheric Cherenkov Telescope (IACT); (2) EAS arrays.

While IACTs are usually more sensitive they run on a reduced duty cycle (15-20 \% of optimal observational conditions) and present a small field of view (around 3-5 degrees). The EAS technique is less sensitive than IACT for the same period of time. However, EAS experiments can offer almost 100 \% duty cycle and a large field of view ($\approx$2 sr, i.e., about 200 times larger than IACTs). 
Therefore, EAS arrays are survey instruments able to monitor continuously a large fraction of the overhead sky. 

Since flaring and bursting events are unpredictable, both in time and location in the sky transient searches can be carried out through the continuous observation of a large fraction of the sky carried out with EAS arrays in the sub-TeV/TeV energy range, and Fermi LAT at GeV energies. While the Fermi-LAT is optimized to observe GeV gamma rays, its sensitivity above a few hundred GeV is mainly limited by the area of the instrument ($\approx$m$^2$), extensive air shower experiments such as HAWC have a much larger effective area at energies above several TeV.

The high-energy gamma rays produced in extragalactic transients are attenuated on their way to Earth by e$^+$e$^-$ pair production on the extragalactic background light (EBL) \cite{lipari}. The attenuation is energy-dependent: as the energy of the $\gamma$-rays increases, the universe becomes increasingly opaque, so that above few TeV it becomes quite difficult to observe $\gamma$-ray sources even at relatively low redshifts. The attenuation also has the effect of distorting energy spectra from distant sources, since the highest-energy $\gamma$-rays reaching the Earth are scattered more than the low-energy $\gamma$-rays. In particular, transient sources such as GRBs are located at cosmological distances from Earth, and the highest energies of $\gamma$-rays from GRBs are likely to be in the 50 to 300 GeV range, depending on the redshift of the source. For this reason, good low-energy sensitivity of arrays is essential for observing extragalactic sources at high redshifts. 

The main obstacle to use current water Cherenkov detectors such as HAWC as transient search detector is the difficulty in the recognition of the photon initiated showers from the hadron initiated ones at energies below 1 TeV. For this reason in HAWC energy events at several hundreds of GeV are not well understood and are thus excluded from the analysis.

All EAS arrays presently in operation or under installation are located in the Northern hemisphere. A new survey instrument located in the Southern Hemisphere should be a high priority even to monitor the Inner Galaxy and the Galactic Center \cite{sgso,disciascio-ricap18}.

In this contribution we introduce the possibility of improving the low energy sensitivity of survey instruments by equipping a suitable Resistive Plate Chambers carpet, which were proved to be optimal detectors at 100 GeV energies by the ARGO-YBJ Collaboration (see, for example, \cite{disciascio-rev} and references therein), with a Water Cherenkov Detectors mainly to improve the $\gamma$/hadron discrimination.
An EAS detector with high sensitivity between 100 GeV and 1 TeV would be a valuable complementary transient detector in the CTA era.

\section{Ground-based Survey Instruments}

Two types of detectors have been used in the last two decades in ground-based survey instruments: (a) Water Cherenkov Detectors (WCDs); (b) Resistive Plate Chambers (RPCs). 

IACTs and WCDs are both based on collecting the Cherenkov emission induced by the cascade of secondary particles in the electromagnetic shower. In the IACTs the Cherenkov light collected is the one produced in the atmosphere, in the WCDs the Cherenkov light is produced in water inside the detectors. This latter is the approach used by Milagro, HAWC and LHAASO. 

With RPCs (the experimental approach used by ARGO-YBJ) the secondary charged particles are detected in the gas gap of the detector.
The benefits in the use of RPCs in ARGO-YBJ, with 2 different read outs, are \cite{disciascio-rev}: (1) high efficiency detection of low energy showers (energy threshold $\sim$300 GeV at 4300 m asl with gamma detection efficiency at 100 GeV of $\sim$50\%) by means of the high density sampling of the central carpet ($\sim$92\% coverage); (2) unprecedented wide energy range investigated by means of the digital/charge read-outs ($\sim$300 GeV $\to$ 10 PeV, with a linearity of the read-out up to $\approx$10$^4$ p/m$^2$); (3) very good energy resolution: about 10\% at 10 TeV and $\sim$ 5\% at 50 TeV for internal events; (4) good angular resolution ($\sigma_{\theta}\approx 1.66^{\circ}$ at the threshold, without any lead layer on top of the RPCs) and unprecedented details in the shower core region by means of the high granularity of the different read-outs.

RPCs allowed to study also charged CR physics (energy spectrum, elemental composition and anisotropy) up to about 10 PeV.
By contrast, the capability of water Cherenkov facilities in extending the energy range to PeV and in selecting primary masses must be still demonstrated.

The key parameters to improve the sensitivity are \cite{disciascio-icrc2017}:
\begin{itemize}
\item \emph{Energy threshold.} Lowering the energy threshold is one of the key to improve the sensitivity and to study flaring phenomena. The low energy threshold of ARGO-YBJ ($\approx$300 GeV) allowed the observation of a large number of extragalactic flares over 4.5 years of monitoring of the Northern sky \cite{argo-mrk421}.
The energy threshold depends on (a) altitude; (b) detection technique and layout; (c) coverage and granularity of the read out; (d) trigger logic; (e) capability to detect the more numerous secondary photons. 
\item \emph{Energy resolution.} The energy resolution of a EAS array is the folding of \emph{Shower fluctuations} and \emph{Sampling fluctuations}. The first term dominates the energy resolution and mainly depends on the depth of the first interaction point. The second term takes into account the fluctuations in the measured number of secondary particles and can be reduced when all particles are measured with a full coverage approach. 
Events with the reconstructed core well inside the instrumented area improve the resolution. 
In the search for PeVatrons a very good energy resolution in the 50 TeV range, where the cut-offs are expected, is needed.
\item \emph{Angular resolution.} The angular resolution depends on the shower multiplicity, zenith angle and extension of the lever arm in the fitting procedure of the temporal profile. The resolution can be improved even by a factor 1.5x - 1.8x at the threshold exploiting the so-called \emph{"Rossi transition effect"}, i.e. by absorbing the low energy delayed particles and by converting the  \emph{'faster'} secondary photons, 
with 1 r.l. of lead or a suitable layer of water above the detectors \cite{argo-angulresol}.
\item \emph{Effective Area.} The effective area is function of the number of charged particles, dimension and coverage of the detector, trigger logic.
\item \emph{Gamma/Hadron discrimination.} This is the key problem of ground-based detectors. The $\gamma$/hadron discrimination below the TeV is still an open issue for EAS arrays and new approaches should be investigated as, for example, a joint analysis of temporal and spatial characteristics of showers. 

The classical method to select the events induced by the background of charged cosmic rays with arrays is to look for \emph{"muon poor"} showers. To evaluate the power of this background rejection technique it is important to know how frequently hadronic showers fluctuate in such a way to have a low muon content as the one resulting from $\gamma$-induced events \cite{disciascio-mu}. The main limitations of this technique is due to the extent of fluctuations in hadron-initiated showers and to the small number of muons. A very large muon detector is needed to reduce the sampling fluctuations. But the small number of muons (only $\approx$3 muons inside a 150$\times$150 m$^2$ area around the shower core is expected in 1 TeV proton-induced showers) make this technique effective only for energies greater than a few TeV.

Hadronic showers typically deposit large amounts of energy in distinct clumps far from the shower core ($>$40 m). Therefore, background rejection can be exploited using topological cuts in the hit pattern, i.e., in the pattern of energy deposition in the detector, of events with the core reconstructed well inside the instrumented area. This is the approach used by WCDs like HAWC and LHAASO. In this case dimensions of the detectors matter. A limitation is that to study topology a minimum number of hits is needed for an optimal discrimination, i.e. this technique is effective above the TeV range. 

The small dimension of detectors explain the poor $\gamma$/hadron discrimination in Milagro and ARGO-YBJ. 
Although some improvements in the angular resolution have been obtained by ARGO-YBJ exploiting suitable cuts in the fitting procedure of the temporal profile, at the threshold the increased statistics with all events resulted in a better sensitivity.
In fact, in ARGO-YBJ the high granularity of the read out allowed, for the first time, a good reconstruction even of events with the core outside the instrumented area.
\end{itemize}

Some characteristics of these different techniques are summarized in the Table 1 \cite{hawc,hawc-crab,argo-crab,argo-spt}.

\begin{table}[h]
\footnotesize{
\caption{\label{tab:one-prime} Main characteristics of WCD- and RPC-based experiments}
\begin{center}
\begin{tabular}{|c|c|c|c|c|c|c|}
\hline
 Experiment &Altitude & Instrumented & Coverage & Energy                & Energy & Angular Resol.\\
                   & (m)       & Area (m$^2$) &                 & Threshold (Crab) & Resol. & (1 $\sigma$ at 1 TeV)\\
  \hline
HAWC (WCD) & 4110 & 140$\times$140 & 60\% & 1 TeV & 55\% - 30\% (1 - 50 TeV) & 0.5$^{\circ}$ \\
ARGO-YBJ (RPC) & 4300 & 100$\times$110 & 92\% & 340 GeV & 10\% - 5\% (10 - 50 TeV)& 0.8$^{\circ}$ \\
\hline
\end{tabular} 
\end{center}
}
\end{table} 
%

Comparison is not straigthforward mainly due to the different dimensions of the detectors. Anyway, WCDs cannot compete with a high granularity carpet of RPCs for what concern the energy threshold and the energy resolution for internal events. 
The energy threshold refers to the median energy of the first multiplicity bin used in the Crab Nebula analysis. We note that in ARGO-YBJ the detection efficiency of 100 GeV $\gamma$-induced events was about 50\%.
The angular resolution, as mentioned, depends mainly on the extension of the lever arm and on the capability to improve the quality of the temporal profile exploiting the Rossi transition effect through a lead layer or a suitable amount of water above the detectors. In the ARGO-YBJ, unfortunately, the lead layer was never installed.

The single counting rate of ARGO-YBJ pads at 4300 m asl is about 380 Hz. The counting rate of 8'' (10'') PMTs in HAWC is 20-30 kHz (40-50 kHz) \cite{hawc-crab}. 
This explains why water Cherenkov facilities will hardly be able to lower the energy threshold well below $\sim$1 TeV. 


\section{The STACEX proposal}

The key to lower the energy threshold in the 100 GeV range is to locate the detector at extreme altitude (about 5000 m asl). But the energy threshold, as well as the angular resolution, as mentioned, depends also on the detection technique, on the coverage (the ratio between the detection area and the instrumented one), on the granularity of the read-out and on the trigger logic.

The aim of the STACEX proposal is to combine in a hybrid detector the two different experimental techniques operated for many years at high altitude:
\begin{itemize}
\item[(a)] a 150$\times$150 m$^2$ RPC carpet, with a 0.5 mm lead layer above, mainly to have
\begin{itemize}
\item dense sampling for a very low energy threshold ($\sim$ 100 GeV);
\item wide energy range: $\sim$100 GeV $\to$ $\sim$10 PeV, to open the PeV range to $\gamma$-ray astronomy and to study classical CR physics;
\item high granularity of the read-out to have good angular resolution, good energy resolution and unprecedented details in the sampling of the temporal profile to improve the background rejection below the TeV.
\end{itemize}
\item[(b)] Water Cherenkov detector below the carpet, mainly to exploit the gamma/hadron discrimination above the TeV through muon detection;
\end{itemize}

\begin{figure}[ht]
\centering
\includegraphics[scale=0.70]{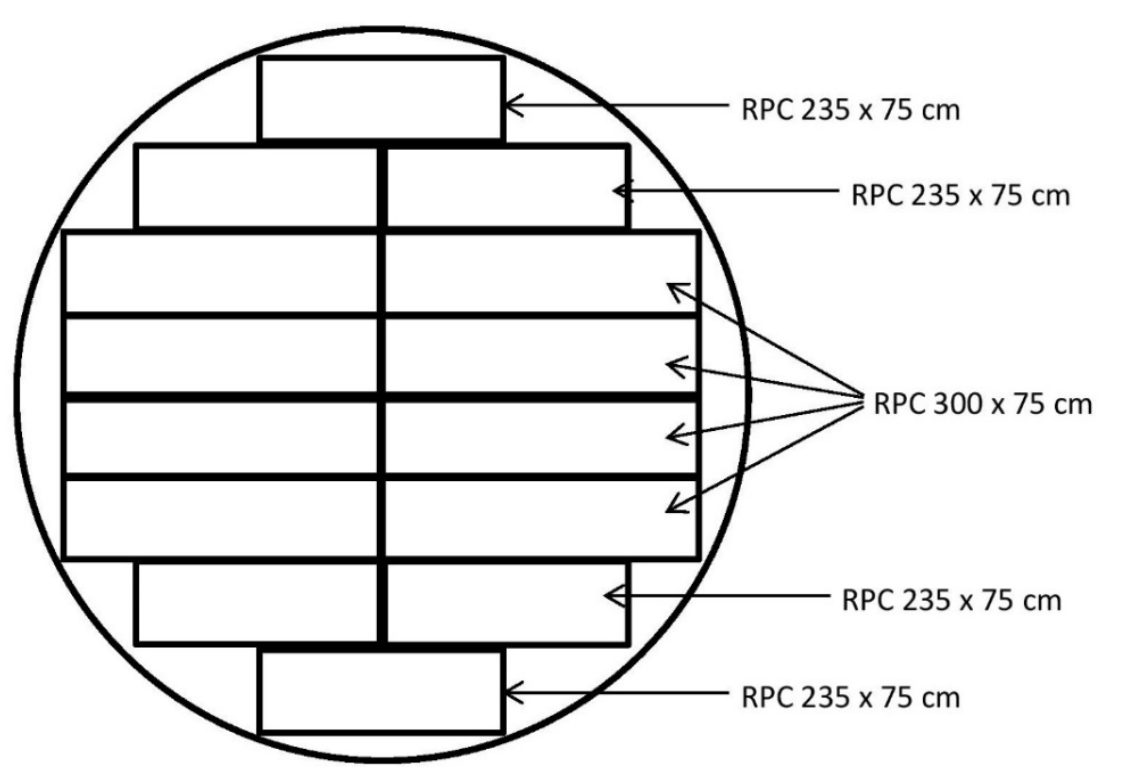}
\caption{Layout of a carpet of RPCs covering a HAWC-like tank surface.}
\label{fig:hybrid}       
\end{figure}

As an example, for a high-altitude test with a HAWC-like tank, in Fig. \ref{fig:hybrid} a possible layout of a carpet of RPCs inside the tank is shown.
The RPCs, located close to the water surface, above the bag containing the water, are mechanically supported by a light mechanical support.
The RPC rectangular geometry is adapted to the circular geometry of the tank, with 7 m diameter, by means of two different chamber sizes. 
There are 8 chambers of size 3$\times$0.75 m$^2$ and 6 chambers of 2.35$\times$0.75m$^2$. The total coverage is about 28 m$^2$ (about 74\% of the HAWC tank surface).

%

\subsection{RPC Detectors}

The type of RPC detector foreseen for this proposal is similar to the well tested ARGO-YBJ chambers, with a few differences suggested by the evolution of the detector in the last 10 - 15 years:
\begin{itemize}
\item operation in avalanche mode instead of streamer;
\item thinner electrode plates: 1.2 mm instead of 2 mm;
\item new front-end electronics, adequate to the avalanche mode operation.
\end{itemize}
%
\begin{figure}[ht]
\centering
\includegraphics[scale=0.70]{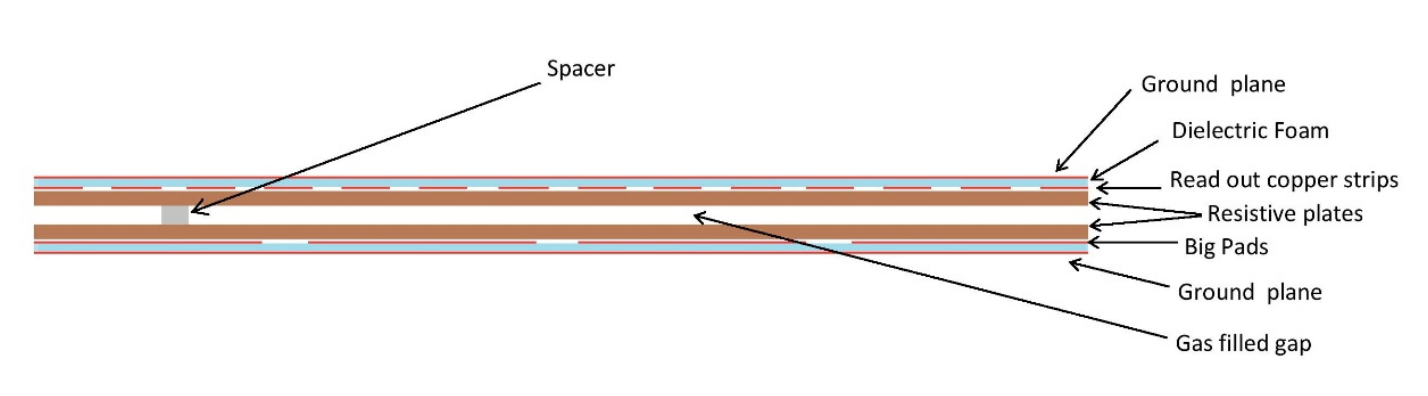}
\caption{RPC layout.}
\label{fig:rpc}       
\end{figure}
%

Each chamber consists of 3 parts assembled together: (1) the gas volume (or gas gap); (2) the strip panels for the signal read out, with the front-end electronics embodied in them; (3) the mechanical support panels. The detector layout is shown in Fig. \ref{fig:rpc}.

\begin{figure}[ht]
\centering
\includegraphics[scale=0.70]{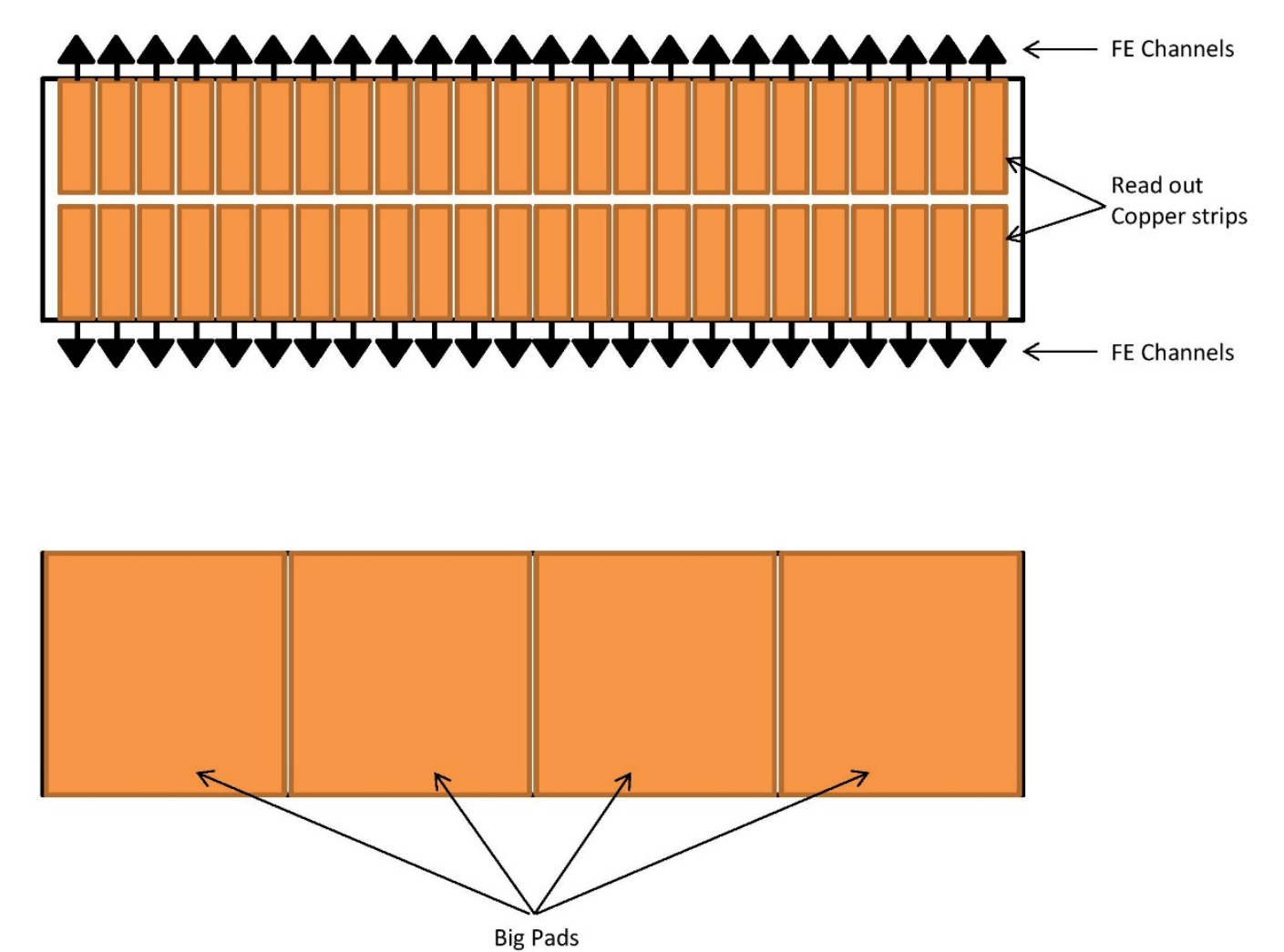}
\caption{The structure of the two read out panels.}
\label{fig:rpc2}       
\end{figure}

The gas volume is a 2 mm gap between two resistive electrode plates each 1.2 mm thick (instead of 2 mm as in ARGO-YBJ). These electrodes are made of high-pressure laminate (HPL) based on phenolic resin, as in ARGO-YBJ. 
The gas volume is sandwiched between two read out panels, which detect the signals generated inside the gas gap. 
The top one is equipped with strips of size 6$\times$40 cm$^2$, with the front-end electronics located at the end of the strip along the chamber edges, as shown in Fig. \ref{fig:rpc2}. 
Each electronics channel consists of an amplifier followed by a discriminator, which supplies a fixed amplitude output signal suitable for timing purposes. The total number of strips is 100 for the 3 m chambers and 78 for the 2.35 m ones. The overall number of strips is 1268. The front-end electronics is embodied inside the detector Faraday cage, as for ARGO-YBJ. 
%
%

This strip panel provides the \emph{"digital"} read out (output = 1 or 0 for each strip) which gives the number of hits of the shower and saturates when the shower hit density approaches or exceeds the number of strips per unit area. 
In order to extend the energy range of the detector well above the 100 TeV domain, the chamber can also be equipped with the \emph{"analog"} read out, which is provided by the bottom panel in Fig. \ref{fig:rpc2}. This consists of 4 (or 3 for the short chambers) \emph{"big pads"}, working as large capacitances integrating the charge induced by all hits falling in the pad. The amplitude reached by this analog signal allows to measure the shower hit density up to about 2$\times$10$^{4}$ m$^{-2}$ in the streamer working mode and even more in avalanche mode.

Finally, the set of the gas volume and the two read out panels are pressed together and mechanically supported by two light honeycomb structures, which provide the stiffness needed for a self-supporting chamber.
In order to reduce the number of TDC channels required for the time readout, 8 (or 16) contiguous strips are grouped in a \emph{"pad"} (like in ARGO-YBJ). Each pad is connected to a TDC channel. The pad is therefore the timing unit to be used in the trigger logic. For each triggered event, the fired strips in each pad are identified off line.

\section{Conclusions}
In the next decade CTA-North and LHAASO \cite{lhaaso} are expected to be the most sensitive detectors to study gamma-ray astronomy in the Northern hemisphere from about 20 GeV up to PeV.

A survey instrument to monitor the Inner Galaxy and the Galactic Center with high sensitivity should be a high priority.
Extreme altitude ($\sim$4500 m asl), high coverage coupled to a high granularity of the read-out are the key to lower the energy threshold at the 100 GeV level and improve the sensitivity to gamma-ray sources. A detector able to sample shower up to particle densities of $\approx$10$^4$ - 10$^5$ p/m$^2$ will allow to study primary cosmic rays up to $\sim$10 PeV.

The ARGO-YBJ Collaboration demonstrated that RPCs can be safely operated at extreme altitudes for many years. 
The benefits in the use of RPCs in ARGO-YBJ were: (1) high efficiency detection of low energy showers (energy threshold $\sim$300 GeV) by means of the dense sampling of the central carpet; (2) unprecedented wide energy range investigated by means of the digital/charge read-outs ($\sim$300 GeV $\to$ 10 PeV); (3) good energy and angular resolutions with unprecedented details in the shower core region by means of the high granularity of the read-outs.
The ARGO-like RPCs should be an important element of a future experiment in the South, possibly coupled to a water Cherenkov detector to exploit the added values of these two experimental approaches. 

The aim of the STACEX proposal is to exploit the capability of a suitable RPC carpet to have a 100 GeV energy threshold, as demonstrated by the ARGO-YBJ Collaboration, coupled with a WCD located below the carpet, mainly devoted to the detection of muons for $\gamma$/hadron discrimination.

\end{document}